\documentclass[%
 reprint,
 amsmath,
 amssymb,
 aps,
 longbibliography,
]{revtex4-1}
\usepackage{graphicx}
\usepackage{dcolumn}
\usepackage{bm}

\begin{document}

\preprint{APS/123-QED}

\title{Propagation mechanism of localized wave packet in plane-Poiseuille flow}

\author{Yue Xiao}
\author{Jianjun Tao}%
 \email{jjtao@pku.edu.cn}
\author{Linsen Zhang}
\affiliation{%
CAPT-HEDPS, SKLTCS, Department of Mechanics and Engineering science, College of Engineering, Peking University, Beijing, 100871 P. R. China
}%




\date{\today}

\begin{abstract}
The convection velocity of localized wave packet in plane-Poiseuille flow is found to be determined by a solitary wave at the centerline of a downstream vortex dipole in its mean field after deducting the basic flow. The fluctuation component following the vortex dipole oscillates with a global frequency selected by the upstream marginal absolute instability, and propagates obeying the local dispersion relation of the mean flow. By applying localized initial disturbances, a nonzero wave-packet density is achieved at the threshold state, suggesting a first order transition.

\end{abstract}

\maketitle

Localized turbulent structures are revealed recently to be the key features near the onset of turbulence in linearly stable shear flows, e.g. puffs in pipe flow and oblique turbulent stripes or bands in channel flows \cite{Eckhardt07, Tuckerman20}. For two-dimensional (2D) plane-Poiseuille flow, the corresponding structure is localized wave packet (LWP) \cite{Rozhdestvensky84, Jimenez90, Price93}, whose relations with finite-amplitude periodic waves were analyzed theoretically and numerically \cite{Soibelman91,Drissi99, Mellibovsky15}. LWP has a strong downstream edge and a slowly decaying upstream edge \cite{Jimenez90}, and the corresponding decay and growth rates were explained in terms of the linear spatial modes \cite{Barnett17}. Similar asymmetry between the upstream and downstream sides was also found for three-dimensional coherent structures in channel flows \cite{Zammert16}. LWP in linearly unstable channel flows, where the Reynolds numbers are larger than the linear critical value 5772, is studied as well, but a damping filter is used in the simulations to restrain LWP from expanding \cite{Teramura16}. The crucial questions for LWP during the subcritical transition are the following: What is the localization mechanism? What determines its convection velocity? What is the selection criterion for the dominating frequency? If the wavelength is not uniform in the streamwise direction, what is its selection criterion?

It is postulated that the subcritical transitions of shear flows may fall into the universality class of directed percolation (DP) \cite{Pomeau86, Sipos11, Shih16, Lemoult16, Pomeau16, Chantry17}. For plane-Poiseuille flow, recent experiments defined a critical Reynolds number $Re_c$ of 830 based on the DP power law \cite{Sano16}, while numerical simulations revealed that the DP power law is retrieved only as $Re$ is above 924 \cite{Shimizu19}. When $Re$ is far below these DP thresholds, it has been found numerically and experimentally that the localized turbulent bands can extend obliquely \cite{Xiong15, Kanazawa17, Paranjape19, Xiao20}, and the periodic turbulent bands can sustain in a sparse turbulent state \cite{Tao18}. By applying random initial disturbances, LWP density, the corresponding parameter of turbulence fraction for two dimensional plane-Poiseuille flow, was shown numerically to approach zero as $Re_c$ was approached from above, and it was concluded that the subcritical transition was more like a continuous phase transition rather than a first-order one \cite{Wang15}. It is known that the subcritical transition may start at different ${Re_c}'s$ depending on different initial or upstream disturbances \cite{Mullin11}. Are the random disturbances the most effective perturbation to trigger the transition at the lowest $Re_c$? Finding answers to these crucial questions is the motivation of this paper.

\begin{figure}[htp]
\includegraphics[width=0.45\textwidth]{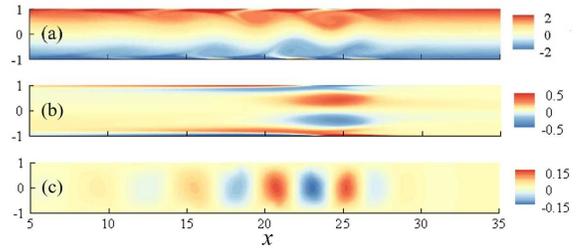}

\caption{\label{fig1}
Contours of (a) the transient vorticity field, (b) the vorticity of the mean-flow modification $\textbf{U}_1$, and (c) the transient normal fluctuation velocity $v'$ obtained numerically in a frame moving with a velocity $c_p = 0.69$ at $Re=2400$. The computational domain is 100 units long.
}
\end{figure}
The incompressible two-dimensional Navier-Stokes equations are solved for plane-Poiseuille flow with a spectral code \cite{Chevalier2007}. The half height of the channel $h$ and 1.5 times of the bulk velocity $U_m$ are chosen as the characteristic length and velocity, respectively. The flow rate is kept constant and the Reynolds number is defined as $Re=1.5U_m h/\nu$, where $\nu$ is the kinematic viscosity of the fluid. Periodic boundary conditions are used in the streamwise $x$ direction and no-slip conditions are imposed at the parallel walls ($y=\pm1$). 65 Chebyshev modes in $y$ direction and 512 Fourier modes per 100 length units in $x$ direction are used. For details of the simulation methods, we refer to the previous papers \cite{Xiong15, Tao18}.

Following the method proposed for turbulent bands \cite{Tao18}, the center's coordinate $x_c$ and length $l$ of the localized wave packet at a given time are defined as,
\begin{equation}
x_c=\frac{\int e x dxdy}{\int edxdy}, \
l=\sqrt{12[\frac{\int e x^2 dxdy}{\int edxdy} -(
\frac{\int ex dxdy }{\int edxdy})^2]},
\end{equation}
where $e$ is the kinetic energy of the velocity field after deducting the basic flow. The packet velocity $c_p$ can be calculated by tracking $x_c$, e.g., $c_p=0.69$ is obtained based on the $x_c$ data during 2000 time units for an isolated LWP at $Re=2400$. It is found that $l$ is large and increases with $Re$ at moderate Reynolds numbers, e.g., $l$ increases from 16.3 at $Re=2500$ to 33.1 at $Re=4500$. In order to diminish the influences of periodic boundaries and obtain the intrinsic properties of an isolated LWP, a computational domain of at least 3$l$ long is required in simulations.

In a frame moving with the packet velocity $c_p$, which is referred to as $S$ frame hereafter, the envelope of LWP looks static and the velocity field is decomposed into three parts,
\begin{equation}
 \textbf{u}(x,y,t) = \textbf{U}_0(y) + \textbf{U}_1(x,y) + \textbf{u}'(x,y,t),
\end{equation}
 where the basic flow $U_0=1-y^2-c_p$, $\textbf{U}_1=(U_1, V_1)$ is the mean-flow modification, the mean flow after deducting $\textbf{U}_0$, and $\textbf{u}'=(u',v')$ is the fluctuation velocity. From now on, $x$ and $y$ denote the coordinates in the $S$ frame. As shown in Fig.\ref{fig1}(a), two rows of vortices lie near the top and bottom walls respectively, a typical feature of LWP \cite{Jimenez90}. The fluctuation component $\textbf{u}'$ travels upstream in a wave form in the $S$ frame [Fig. 1(c)], and is referred to as fluctuation wave (FW) in this paper. The mean flow in $S$ frame is calculated with 200 fields sampled every 10 time units. It is illustrated in Fig. 1(b) that the vorticity field of the mean-flow modification exhibits a prominent structure: a vortex dipole at the downstream end of LWP sandwiched between vortex layers extending upstream near the walls. The vortex dipole, to the best of our knowledge, has not been reported before, and is shown next to be responsible for the localization property, the packet velocity $c_p$, and the frequency of LWP.

\begin{figure}
\includegraphics[width=0.47\textwidth]{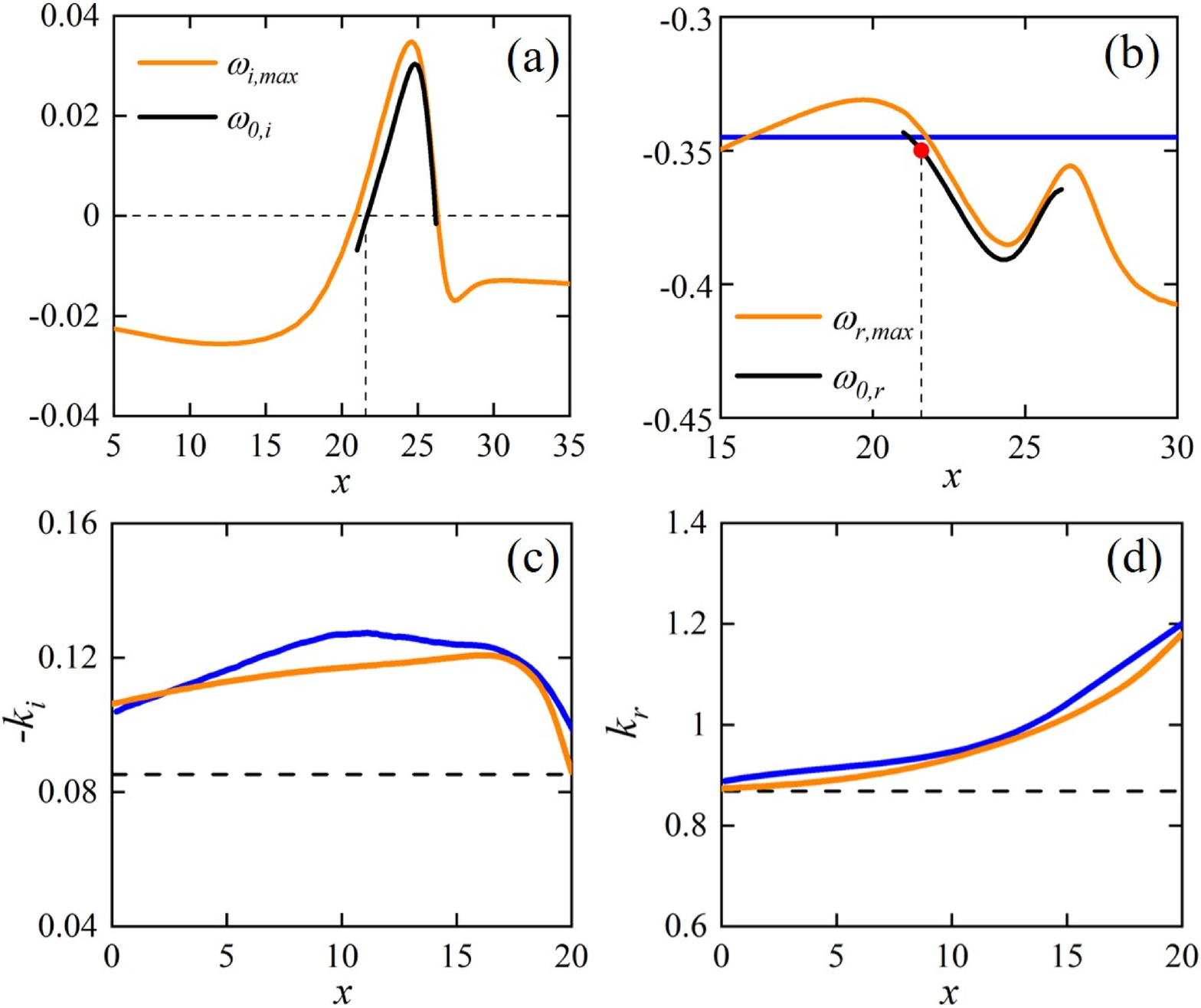}
\includegraphics[width=0.45\textwidth]{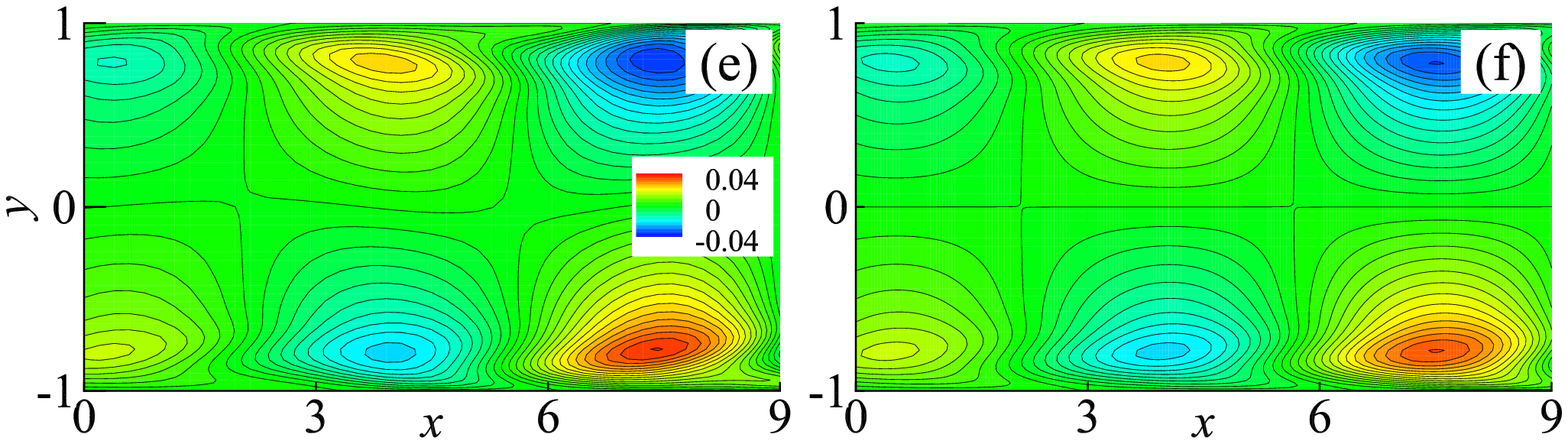}
 \caption{\label{fig2}%
The maximum temporal growth rate $\omega_{i,max}$ and its corresponding frequency $\omega_{r,max}$ of temporal mode, and the absolute growth rate $\omega_{0,i}$ and absolute frequency $\omega_{0,r}$ at different $x$ are shown in (a) and (b), respectively. The vertical dashed lines indicate the upstream position of marginal absolute instability. The minimal $-k_i$ and the corresponding $k_r$ of spatial mode with the global frequency $\omega_g = -0.345$ are shown by orange curves in (c) and (d), respectively. The oscillating frequency, wavenumber, and spatial growth rate of FW obtained in simulations are shown as blue lines in (b), (c), and (d), respectively. The corresponding parameters solved for the basic flow $\textbf{U}_0$ are shown as horizontal dashed lines in (c) and (d), respectively, for reference. The iso-contours of FW's $u'$ and of the streamwise disturbance velocity of spatial mode solved based on the mean-flow profile at $x=5$ are shown in (e) and (f), respectively. $Re$=2400. }%
\end{figure}

 In order to understand the dynamic behavior of FW, linear stability analyses are carried out for the mean flow ($\textbf{U}_0+\textbf{U}_1$) corresponding to Fig. 1(b) at different $x$ based on the parallel-flow approximation. The perturbations are assumed in the form of $\sim e^{i[(k_r+ik_i)x-(\omega_r+i\omega_i)t]}$, where $k_r$, $k_i$, $\omega_r$, and $\omega_i$ are solved from the local dispersion relation \cite{Schmid01} and shown in Fig. 2. Several interesting features should be noted. First, there is a finite unstable region ($\omega_{i,max}>0$) surrounded by stable regions, i.e. the region between $x=21$ and 26.3 shown in Fig. 2(a), a main part of the vortex dipole region [Fig. 1(b)]. Second, $\omega_{r,max}$, the frequency of the most unstable temporal mode ($k_r>0,\ k_i=0$) is negative as shown in Fig. 2(b), representing an unstable wave mode traveling upstream in the $S$ frame just as the FW found in simulations. This result is reasonable by considering that the mode's maximum amplitude lies at the neighborhoods of walls [Fig. 2(f)], where the mean flow moves upstream. These  features suggest a localization mechanism for LWP: traveling wave mode amplified in the unstable region decays in the stable regions, forming a localized wave packet.

  According to the simulations, FW propagates with a unique global frequency, e.g., $\omega_g = -0.345$ as shown by the blue line in Fig. 2(b). It is noted that though the frequency can be measured as $x>30$, the fluctuation velocity is too weak to be recognized as shown in Fig. 1(c) because the downstream side of the present LWP is the far upstream end of another LWP. In order to understand the selection criterion of the global frequency, spatio-temporal stability analyses are carried out based on the mean flow, and the absolute growth rate $\omega_{0,i}$ and absolute frequency $\omega_{0,r}$, where the group velocity is zero, are shown in Fig. 2.  $\omega_{0,r}$ satisfying the saddle-point condition \cite{Chomaz91, Hammond97} is computed as -0.39, which is different from $\omega_g$, and hence the saddle-point criterion seems not applicable for LWP. The absolute frequency at the upstream boundary of the absolutely unstable region ($\omega_{0,i}>0$) is -0.35, which is labeled by the red point in Fig. 2(b) and almost coincides with $\omega_g$, suggesting an  marginal stability criterion for the global frequency selection \cite{Dee83, Monkewitz87}.

Based on the fluctuation vorticity recorded during 2000 time units along the bottom wall, the phase velocity and then the wave number of FW are determined with $\omega_g$ at each $x$ position, and the spatial growth rate of FW is calculated based on the the envelope of the normal fluctuation velocity $v'$ along the centerline. Since the wave following the dipole decays in the upstream direction, we look for the spatial mode with the minimal $-k_i$ at each $x$. As shown in Fig. 2(c) and 2(d), the spatial growth rate and wave number of FW acceptably agree with $-k_i$ and $k_r$ of the spatial mode with $\omega_g$, and the flow structure of FW shown in Fig. 2(e) agrees with that of the spatial mode [Fig. 2(f)] as well, indicating that the spatial properties of FW are mainly determined by the local dispersion relation of the mean flow.

The spatial instability of LWP was discussed before based on the basic flow and the frequency obtained in simulations, and constant $k_r$ and $k_i$ were solved and found to be consistent with the simulations at upstream and downstream tails of LWP \cite{Barnett17}. Such a consistency is an asymptotic case for the present study. As shown in Fig. 2, the spatial growth rate and wave number of FW are not constant but vary in the streamwise direction. However, at the far upstream tail of LWP, where the mean flow modification almost diminishes, $-k_i$ and $k_r$ are close to the asymptotic values corresponding to the basic flow as shown by the horizontal dashed lines in Fig. 2(c) and 2(d), respectively.

\begin{figure}[htb]
\includegraphics[width=0.31\textwidth]{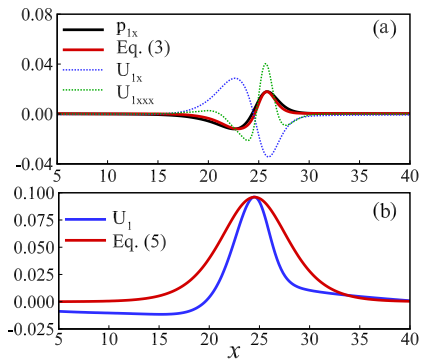}
\caption{\label{fig3}
    (a) Derivatives of the mean pressure modification $P_1$ and the mean velocity modification $U_1$ along the centerline shown in Fig. 1(b), and (b) $U_1$ compared with the solitary-wave solution Eq. (5).
}
\end{figure}

Besides FW, another key feature of LWP is its propagation velocity.  As shown in Fig. 1(b), the vorticity field is antisymmetric about the centerline, and hence at the centerline $V_1$ and $\partial U_1/\partial y$ or $U_{1y}$ are zero. Consequently, the mean flow along the centerline may be described by a steady one-dimensional nonlinear model. Considering that the subcritical transitions occur as $1/Re\sim 10^{-3}$ and the streamwise fluctuation velocity $u'$ is very small along the centerline as shown in Fig.2(e), the viscous diffusion of the mean flow modification and the Reynolds stress ($\overline{u'v'}$ and $\overline{u'u'}$) are ignored along the centerline. As a result, the gradient of the mean pressure modification $P_{1x}$ should mainly depend on the variation and the derivatives of the centerline mean velocity, i.e. $(1-c_p+U_1)$, $U_{1x}$, $U_{1xx}$, $U_{1xxx}$, .... According to the Bernoulli equation, deceleration corresponds to an adverse pressure gradient, suggesting that $P_{1x}$ remains roughly the opposed phase to $U_{1x}$, an odd derivative of $U_1$, as shown in Fig. 3(a). Inspired by the fact that even derivatives will cause phase shift from the odd derivatives for a harmonic wave, only the odd derivatives of $U_1$ are considered and the dependency of $P_{1x}$ on $U_1$ is simplified to a linear relation:
\begin{equation}
 P_{1x} =A U_{1x} +B U_{1xxx},
\end{equation}
where $A$ and $B$ are coefficients.  As shown in Fig. 3(a), the pressure gradient estimated with the velocity derivatives (red curve) agrees with the numerical data (black curve) along the centerline of the vortex dipole, indicating that Eq. (3) grasps the main relation between $P_x$ and $U_1$. Note that $A=-0.347$ and $B=0.154$ are used in Eq. (3) to guarantee that the estimated $P_{1x}$ has the same minimum and maximum values as the numerical one.

Substituting Eq. (3) into the mean $x-$momentum equation, the steady centerline model for the mean flow becomes:
\begin{equation}
\left\{
\begin{array}{c}
(1+A-c_p+U_1)U_{1x}+BU_{1xxx}=0 \\
U_1(\infty)=U_{1x}(\infty)=U_{1xx}(\infty)=0
\end{array}
\right.
\end{equation}
This is a KdV-type equation for a steady solitary wave in the moving $S$ frame or equivalently, a solitary wave traveling with a velocity of $c_p$ in a motionless frame. Its solution can be solved easily as:
\begin{equation}
U_1=3(c_p-1-A)sech^2[\sqrt{\frac{c_p-A-1}{4B}}(x-x_1)],
\end{equation}
where $x_1$ is a constant defining the $x$ coordinate of the maximum $U_1$. As shown in Fig. 3(b), the maximum velocity predicted by the centerline model [Eq. (5)] is consistent with the corresponding simulation value. Therefore, when the maximum mean velocity at the centerline of the vortex dipole region is given, $c_p$, the convection velocity of the solitary wave and LWP in a motionless frame, is determined by Eq. (5). Interestingly, though $A$ and $B$ used in Fig. 3 are determined with the numerical data for $Re=2400$, the nonlinear centerline model [Eq. (3)-(5)] does not include the Reynolds number explicitly, suggesting that $c_p$ does not strongly depend on $Re$. This suggestion agrees qualitatively with the numerical simulations, where the convection velocity of LWP increases slightly with $Re$, e.g. $c_p=0.69$, 0.74, and 0.78 for $Re=2400$, 3000, and 4000, respectively.

It should be noted that without the Reynolds stress contributed by the finite-amplitude FW the vortex dipole will decay due to viscous diffusion and dissipation. It is checked numerically that both the finite-amplitude $\textbf{U}_1$ and the finite-amplitude FW are necessary to sustain LWP, denoting that there exists a threshold energy and the subcritical transition is a first order transition. LWP becomes longer with the increase of $Re$, and was found to split at $Re=5000$ \cite{Jimenez90}. According to the present simulations, the critical Reynolds number for an isolated LWP to split in a long domain is about 4950, above which LWP splitting will cause more LWPs and the whole domain will be occupied by LWPs eventually, leading to a  statistically steady or equilibrium state.

\begin{figure}[htb]
\includegraphics[width=0.52\textwidth]{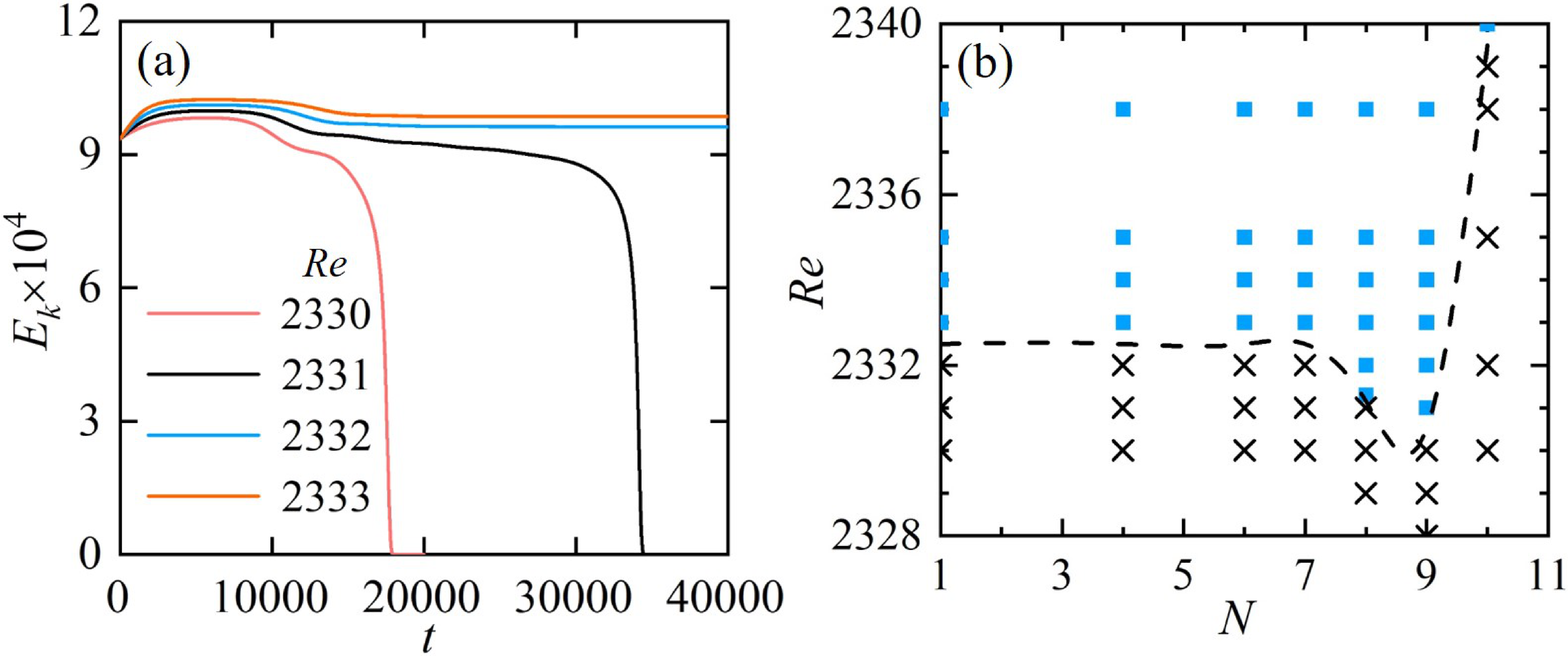}

\caption{\label{fig4}
(a) Time series of the perturbation kinetic energy $E_k$ at different Reynolds numbers with the same initial field containing 8 LWPs. (b) The squares and crosses represent the cases where the initially introduced $N$ LWPs are sustained and vanish within 40000 time units, respectively. The length of computational domain is 400.
}
\end{figure}

When we decrease $Re$ gradually from the equilibrium state, e.g. $Re=6000$, the perturbation kinetic energy in the whole domain area $S$, $E_k =\frac{1}{2S}\int_S \mid \textbf{U}_1 + \textbf{u}'\mid^2 dS$, decreases but does not vanish if $Re$ is larger than 2331, where 9 LWPs are reserved in a domain of 400 units long. In order to examine the $Re$ threshold for sustained LWP, different numbers ($N$) of a sample LWP obtained at $Re=2350$ are evenly spaced and introduced as initial perturbations. As shown in Fig. 4(a), relaminarization occurs when $Re$ becomes lower than a critical value $Re_c$, and the same $Re_c = 2332.5\pm 0.5$ corresponds to different $N's$ [Fig. 4(b)] or LWP densities (the LWP number per unit streamwise length) less than 0.017, indicating that when LWPs are far from each other they behave just like an isolated one. According to the dashed curve shown in Fig. 4(b), the lowest $Re_c$ or threshold $Re$ for sustained LWP is $2330.5\pm 0.5$ with a nonzero threshold LWP density about 0.022, confirming again that the present transition is a discontinuous type. The LWPs initially arranged too tightly (e.g. the case of $N=10$ shown in Fig. 4) tend to decay at low Reynolds numbers due to LWP interaction. Considering that random initial disturbances are not as effective as LWP themselves to trigger LWPs and the LWPs caused by random disturbances through transient growth may stay tightly and tend to diminish, using random initial disturbances may lead to less sustained LWPs than localized initial perturbations.

According to the present study, the localized wave packet found in two-dimensional plane-Poiseuille flow represents a symbiosis between the vortex dipole in the mean flow modification and the fluctuation wave: the dipole defines the convection velocity of the whole packet with the solitary wave velocity at its centerline, provides an unstable region to amplify FW and a global frequency for FW, while FW feeds back the Reynolds stress to prevent the vortex dipole from decaying, and travels upstream obeying the local dispersion relation of the mean flow. Therefore, nonlinear effects are necessary to sustain LWP, and the subcritical transition is a first order transition with a nonzero threshold LWP density. In addition, isolated LWP can sustain as $Re<4950$, suggesting that the initial transition stage is characterized by a sparse structure state instead of an equilibrium state, which can be achieved only as $Re>4950$ due to the wave packet split.

\begin{acknowledgments}
The simulation code SIMSON from KTH and help from P. Schlatter, L. Brandt, and D. Henningson are gratefully acknowledged. The simulations were performed on TianHe-1(A). This work is supported by the National Natural Science Foundation of China (Grants No. 91752203, and No. 11490553).
\end{acknowledgments}



\begin{thebibliography}{32}%
\makeatletter
\providecommand \@ifxundefined [1]{%
 \@ifx{#1\undefined}
}%
\providecommand \@ifnum [1]{%
 \ifnum #1\expandafter \@firstoftwo
 \else \expandafter \@secondoftwo
 \fi
}%
\providecommand \@ifx [1]{%
 \ifx #1\expandafter \@firstoftwo
 \else \expandafter \@secondoftwo
 \fi
}%
\providecommand \natexlab [1]{#1}%
\providecommand \enquote  [1]{``#1''}%
\providecommand \bibnamefont  [1]{#1}%
\providecommand \bibfnamefont [1]{#1}%
\providecommand \citenamefont [1]{#1}%
\providecommand \href@noop [0]{\@secondoftwo}%
\providecommand \href [0]{\begingroup \@sanitize@url \@href}%
\providecommand \@href[1]{\@@startlink{#1}\@@href}%
\providecommand \@@href[1]{\endgroup#1\@@endlink}%
\providecommand \@sanitize@url [0]{\catcode `\\12\catcode `\$12\catcode
  `\&12\catcode `\#12\catcode `\^12\catcode `\_12\catcode `\%12\relax}%
\providecommand \@@startlink[1]{}%
\providecommand \@@endlink[0]{}%
\providecommand \url  [0]{\begingroup\@sanitize@url \@url }%
\providecommand \@url [1]{\endgroup\@href {#1}{\urlprefix }}%
\providecommand \urlprefix  [0]{URL }%
\providecommand \Eprint [0]{\href }%
\providecommand \doibase [0]{http://dx.doi.org/}%
\providecommand \selectlanguage [0]{\@gobble}%
\providecommand \bibinfo  [0]{\@secondoftwo}%
\providecommand \bibfield  [0]{\@secondoftwo}%
\providecommand \translation [1]{[#1]}%
\providecommand \BibitemOpen [0]{}%
\providecommand \bibitemStop [0]{}%
\providecommand \bibitemNoStop [0]{.\EOS\space}%
\providecommand \EOS [0]{\spacefactor3000\relax}%
\providecommand \BibitemShut  [1]{\csname bibitem#1\endcsname}%
\let\auto@bib@innerbib\@empty
\bibitem [{\citenamefont {Eckhardt}\ \emph {et~al.}(2007)\citenamefont
  {Eckhardt}, \citenamefont {Schneider}, \citenamefont {Hof},\ and\
  \citenamefont {Westerweel}}]{Eckhardt07}%
  \BibitemOpen
  \bibfield  {author} {\bibinfo {author} {\bibfnamefont {Bruno}\ \bibnamefont
  {Eckhardt}}, \bibinfo {author} {\bibfnamefont {Tobias~M.}\ \bibnamefont
  {Schneider}}, \bibinfo {author} {\bibfnamefont {Bjorn}\ \bibnamefont {Hof}},
  \ and\ \bibinfo {author} {\bibfnamefont {Jerry}\ \bibnamefont {Westerweel}},\
  }\bibfield  {title} {\enquote {\bibinfo {title} {Turbulence transition in
  pipe flow},}\ }\href@noop {} {\bibfield  {journal} {\bibinfo  {journal}
  {Annu. Rev. Fluid Mech.}\ }\textbf {\bibinfo {volume} {39}},\ \bibinfo
  {pages} {447--468} (\bibinfo {year} {2007})}\BibitemShut {NoStop}%
\bibitem [{\citenamefont {Tuckerman}\ \emph {et~al.}(2020)\citenamefont
  {Tuckerman}, \citenamefont {Chantry},\ and\ \citenamefont
  {Barkley}}]{Tuckerman20}%
  \BibitemOpen
  \bibfield  {author} {\bibinfo {author} {\bibfnamefont {Laurette~S.}\
  \bibnamefont {Tuckerman}}, \bibinfo {author} {\bibfnamefont {Matthew}\
  \bibnamefont {Chantry}}, \ and\ \bibinfo {author} {\bibfnamefont {Dwight}\
  \bibnamefont {Barkley}},\ }\bibfield  {title} {\enquote {\bibinfo {title}
  {Patterns in wall-bounded shear flows},}\ }\href@noop {} {\bibfield
  {journal} {\bibinfo  {journal} {Annu. Rev. Fluid Mech.}\ }\textbf {\bibinfo
  {volume} {52}},\ \bibinfo {pages} {343--367} (\bibinfo {year}
  {2020})}\BibitemShut {NoStop}%
\bibitem [{\citenamefont {Rozhdestvensky}\ and\ \citenamefont
  {Simakin}(1984)}]{Rozhdestvensky84}%
  \BibitemOpen
  \bibfield  {author} {\bibinfo {author} {\bibfnamefont {B.~L.}\ \bibnamefont
  {Rozhdestvensky}}\ and\ \bibinfo {author} {\bibfnamefont {I.N.}\ \bibnamefont
  {Simakin}},\ }\bibfield  {title} {\enquote {\bibinfo {title} {Secondary flows
  in a plane channel : their relationship and comparison with turbulent
  flows},}\ }\href@noop {} {\bibfield  {journal} {\bibinfo  {journal} {J. Fluid
  Mech.}\ }\textbf {\bibinfo {volume} {147}},\ \bibinfo {pages} {261--289}
  (\bibinfo {year} {1984})}\BibitemShut {NoStop}%
\bibitem [{\citenamefont {Jimenez}(1990)}]{Jimenez90}%
  \BibitemOpen
  \bibfield  {author} {\bibinfo {author} {\bibfnamefont {Javier}\ \bibnamefont
  {Jimenez}},\ }\bibfield  {title} {\enquote {\bibinfo {title} {Transition to
  turbulence in two-dimensional poiseuille flow},}\ }\href@noop {} {\bibfield
  {journal} {\bibinfo  {journal} {J. Fluid Mech.}\ }\textbf {\bibinfo {volume}
  {218}},\ \bibinfo {pages} {265--297} (\bibinfo {year} {1990})}\BibitemShut
  {NoStop}%
\bibitem [{\citenamefont {Price}\ \emph {et~al.}(1993)\citenamefont {Price},
  \citenamefont {Brachet},\ and\ \citenamefont {Pomeau}}]{Price93}%
  \BibitemOpen
  \bibfield  {author} {\bibinfo {author} {\bibfnamefont {Tim}\ \bibnamefont
  {Price}}, \bibinfo {author} {\bibfnamefont {Marc}\ \bibnamefont {Brachet}}, \
  and\ \bibinfo {author} {\bibfnamefont {Yves}\ \bibnamefont {Pomeau}},\
  }\bibfield  {title} {\enquote {\bibinfo {title} {Numerical characterization
  of localized solutions in plane poiseuille flow},}\ }\href@noop {} {\bibfield
   {journal} {\bibinfo  {journal} {Phys. Fluids A}\ }\textbf {\bibinfo {volume}
  {5}},\ \bibinfo {pages} {762--764} (\bibinfo {year} {1993})}\BibitemShut
  {NoStop}%
\bibitem [{\citenamefont {Soibelman}\ and\ \citenamefont
  {Meiron}(1991)}]{Soibelman91}%
  \BibitemOpen
  \bibfield  {author} {\bibinfo {author} {\bibfnamefont {Isreal}\ \bibnamefont
  {Soibelman}}\ and\ \bibinfo {author} {\bibfnamefont {D.~I.}\ \bibnamefont
  {Meiron}},\ }\bibfield  {title} {\enquote {\bibinfo {title} {Finite-amplitude
  bifurcations in plane poiseuille flow : two-dimensional hopf bifurcation},}\
  }\href@noop {} {\bibfield  {journal} {\bibinfo  {journal} {J. Fluid Mech.}\
  }\textbf {\bibinfo {volume} {229}},\ \bibinfo {pages} {389--416} (\bibinfo
  {year} {1991})}\BibitemShut {NoStop}%
\bibitem [{\citenamefont {Drissi}\ \emph {et~al.}(1999)\citenamefont {Drissi},
  \citenamefont {Net},\ and\ \citenamefont {Mercader}}]{Drissi99}%
  \BibitemOpen
  \bibfield  {author} {\bibinfo {author} {\bibfnamefont {A.}~\bibnamefont
  {Drissi}}, \bibinfo {author} {\bibfnamefont {M.}~\bibnamefont {Net}}, \ and\
  \bibinfo {author} {\bibfnamefont {I.}~\bibnamefont {Mercader}},\ }\bibfield
  {title} {\enquote {\bibinfo {title} {Subharmonic instabilities of
  {T}ollmien-{S}chlichting waves in two-dimensional poiseuille flow},}\
  }\href@noop {} {\bibfield  {journal} {\bibinfo  {journal} {Phys. Rev. E}\
  }\textbf {\bibinfo {volume} {62}},\ \bibinfo {pages} {1781--1791} (\bibinfo
  {year} {1999})}\BibitemShut {NoStop}%
\bibitem [{\citenamefont {Mellibovsky}\ and\ \citenamefont
  {Meseguer}(2015)}]{Mellibovsky15}%
  \BibitemOpen
  \bibfield  {author} {\bibinfo {author} {\bibfnamefont {Fernando}\
  \bibnamefont {Mellibovsky}}\ and\ \bibinfo {author} {\bibfnamefont {Alvaro}\
  \bibnamefont {Meseguer}},\ }\bibfield  {title} {\enquote {\bibinfo {title} {A
  mechanism for streamwise localisation of nonlinear waves in shear flows},}\
  }\href@noop {} {\bibfield  {journal} {\bibinfo  {journal} {J. Fluid Mech.}\
  }\textbf {\bibinfo {volume} {779}},\ \bibinfo {pages} {R1} (\bibinfo {year}
  {2015})}\BibitemShut {NoStop}%
\bibitem [{\citenamefont {Barnett}\ \emph {et~al.}(2017)\citenamefont
  {Barnett}, \citenamefont {Gurevich},\ and\ \citenamefont
  {Grigoriev}}]{Barnett17}%
  \BibitemOpen
  \bibfield  {author} {\bibinfo {author} {\bibfnamefont {Joshua}\ \bibnamefont
  {Barnett}}, \bibinfo {author} {\bibfnamefont {Daniel~R.}\ \bibnamefont
  {Gurevich}}, \ and\ \bibinfo {author} {\bibfnamefont {Roman~O.}\ \bibnamefont
  {Grigoriev}},\ }\bibfield  {title} {\enquote {\bibinfo {title} {Streamwise
  localization of traveling wave solutions in channel flow},}\ }\href@noop {}
  {\bibfield  {journal} {\bibinfo  {journal} {Phys. Rev. E}\ }\textbf {\bibinfo
  {volume} {95}},\ \bibinfo {pages} {033124} (\bibinfo {year}
  {2017})}\BibitemShut {NoStop}%
\bibitem [{\citenamefont {Zammert}\ and\ \citenamefont
  {Eckhardt}(2016)}]{Zammert16}%
  \BibitemOpen
  \bibfield  {author} {\bibinfo {author} {\bibfnamefont {Stefan}\ \bibnamefont
  {Zammert}}\ and\ \bibinfo {author} {\bibfnamefont {Bruno}\ \bibnamefont
  {Eckhardt}},\ }\bibfield  {title} {\enquote {\bibinfo {title} {Streamwise
  decay of localized states in channel flow},}\ }\href@noop {} {\bibfield
  {journal} {\bibinfo  {journal} {Phys. Rev. E}\ }\textbf {\bibinfo {volume}
  {94}},\ \bibinfo {pages} {041101} (\bibinfo {year} {2016})}\BibitemShut
  {NoStop}%
\bibitem [{\citenamefont {Teramura}\ and\ \citenamefont
  {Toh}(2016)}]{Teramura16}%
  \BibitemOpen
  \bibfield  {author} {\bibinfo {author} {\bibfnamefont {Toshiki}\ \bibnamefont
  {Teramura}}\ and\ \bibinfo {author} {\bibfnamefont {Sadayoshi}\ \bibnamefont
  {Toh}},\ }\bibfield  {title} {\enquote {\bibinfo {title} {Chaotic
  self-sustaining structure embedded in the turbulent-laminar interface},}\
  }\href@noop {} {\bibfield  {journal} {\bibinfo  {journal} {Phys. Rev. E}\
  }\textbf {\bibinfo {volume} {93}},\ \bibinfo {pages} {041101} (\bibinfo
  {year} {2016})}\BibitemShut {NoStop}%
\bibitem [{\citenamefont {Pomeau}(1986)}]{Pomeau86}%
  \BibitemOpen
  \bibfield  {author} {\bibinfo {author} {\bibfnamefont {Y}~\bibnamefont
  {Pomeau}},\ }\bibfield  {title} {\enquote {\bibinfo {title} {Front motion,
  metastability and subcritical bifurcations in hydrodynamics},}\ }\href@noop
  {} {\bibfield  {journal} {\bibinfo  {journal} {Physica D}\ }\textbf {\bibinfo
  {volume} {23}},\ \bibinfo {pages} {3--11} (\bibinfo {year}
  {1986})}\BibitemShut {NoStop}%
\bibitem [{\citenamefont {Sipos}\ and\ \citenamefont
  {Goldenfeld}(2011)}]{Sipos11}%
  \BibitemOpen
  \bibfield  {author} {\bibinfo {author} {\bibfnamefont {Maksim}\ \bibnamefont
  {Sipos}}\ and\ \bibinfo {author} {\bibfnamefont {Nigel}\ \bibnamefont
  {Goldenfeld}},\ }\bibfield  {title} {\enquote {\bibinfo {title} {Directed
  percolation describes lifetime and growth of turbulent puffs and slugs},}\
  }\href@noop {} {\bibfield  {journal} {\bibinfo  {journal} {Phys. Rev. E}\
  }\textbf {\bibinfo {volume} {84}},\ \bibinfo {pages} {035304} (\bibinfo
  {year} {2011})}\BibitemShut {NoStop}%
\bibitem [{\citenamefont {Shih}\ \emph {et~al.}(2016)\citenamefont {Shih},
  \citenamefont {Hsieh},\ and\ \citenamefont {Goldenfeld}}]{Shih16}%
  \BibitemOpen
  \bibfield  {author} {\bibinfo {author} {\bibfnamefont {Hong~Yan}\
  \bibnamefont {Shih}}, \bibinfo {author} {\bibfnamefont {Tsung~Lin}\
  \bibnamefont {Hsieh}}, \ and\ \bibinfo {author} {\bibfnamefont {Nigel}\
  \bibnamefont {Goldenfeld}},\ }\bibfield  {title} {\enquote {\bibinfo {title}
  {Ecological collapse and the emergence of travelling waves at the onset of
  shear turbulence},}\ }\href@noop {} {\bibfield  {journal} {\bibinfo
  {journal} {Nature Phys.}\ }\textbf {\bibinfo {volume} {12}},\ \bibinfo
  {pages} {245--248} (\bibinfo {year} {2016})}\BibitemShut {NoStop}%
\bibitem [{\citenamefont {Lemoult}\ \emph {et~al.}(2016)\citenamefont
  {Lemoult}, \citenamefont {Shi}, \citenamefont {Avila}, \citenamefont
  {Jalikop}, \citenamefont {Avila},\ and\ \citenamefont {Hof}}]{Lemoult16}%
  \BibitemOpen
  \bibfield  {author} {\bibinfo {author} {\bibfnamefont {Gregoire}\
  \bibnamefont {Lemoult}}, \bibinfo {author} {\bibfnamefont {Liang}\
  \bibnamefont {Shi}}, \bibinfo {author} {\bibfnamefont {Kerstin}\ \bibnamefont
  {Avila}}, \bibinfo {author} {\bibfnamefont {Shreyas~V.}\ \bibnamefont
  {Jalikop}}, \bibinfo {author} {\bibfnamefont {Marc}\ \bibnamefont {Avila}}, \
  and\ \bibinfo {author} {\bibfnamefont {Bjorn}\ \bibnamefont {Hof}},\
  }\bibfield  {title} {\enquote {\bibinfo {title} {Directed percolation phase
  transition to sustained turbulence in couette flow},}\ }\href@noop {}
  {\bibfield  {journal} {\bibinfo  {journal} {Nature Phys.}\ }\textbf {\bibinfo
  {volume} {12}},\ \bibinfo {pages} {254--258} (\bibinfo {year}
  {2016})}\BibitemShut {NoStop}%
\bibitem [{\citenamefont {Pomeau}(2016)}]{Pomeau16}%
  \BibitemOpen
  \bibfield  {author} {\bibinfo {author} {\bibfnamefont {Yves}\ \bibnamefont
  {Pomeau}},\ }\bibfield  {title} {\enquote {\bibinfo {title} {The long and
  winding road},}\ }\href@noop {} {\bibfield  {journal} {\bibinfo  {journal}
  {Nature Phys.}\ }\textbf {\bibinfo {volume} {12}},\ \bibinfo {pages}
  {198--199} (\bibinfo {year} {2016})}\BibitemShut {NoStop}%
\bibitem [{\citenamefont {Chantry}\ \emph {et~al.}(2017)\citenamefont
  {Chantry}, \citenamefont {Tuckerman},\ and\ \citenamefont
  {Barkley}}]{Chantry17}%
  \BibitemOpen
  \bibfield  {author} {\bibinfo {author} {\bibfnamefont {Matthew}\ \bibnamefont
  {Chantry}}, \bibinfo {author} {\bibfnamefont {Laurette~S.}\ \bibnamefont
  {Tuckerman}}, \ and\ \bibinfo {author} {\bibfnamefont {Dwight}\ \bibnamefont
  {Barkley}},\ }\bibfield  {title} {\enquote {\bibinfo {title} {Universal
  continuous transition to turbulence in a planar shear flow},}\ }\href@noop {}
  {\bibfield  {journal} {\bibinfo  {journal} {J. Fluid Mech.}\ }\textbf
  {\bibinfo {volume} {824}},\ \bibinfo {pages} {R1} (\bibinfo {year}
  {2017})}\BibitemShut {NoStop}%
\bibitem [{\citenamefont {Sano}\ and\ \citenamefont {Tamai}(2016)}]{Sano16}%
  \BibitemOpen
  \bibfield  {author} {\bibinfo {author} {\bibfnamefont {Masaki}\ \bibnamefont
  {Sano}}\ and\ \bibinfo {author} {\bibfnamefont {Keiichi}\ \bibnamefont
  {Tamai}},\ }\bibfield  {title} {\enquote {\bibinfo {title} {A universal
  transition to turbulence in channel flow},}\ }\href@noop {} {\bibfield
  {journal} {\bibinfo  {journal} {Nature Phys.}\ }\textbf {\bibinfo {volume}
  {12}},\ \bibinfo {pages} {249--253} (\bibinfo {year} {2016})}\BibitemShut
  {NoStop}%
\bibitem [{\citenamefont {Shimizu}\ and\ \citenamefont
  {Manneville}(2019)}]{Shimizu19}%
  \BibitemOpen
  \bibfield  {author} {\bibinfo {author} {\bibfnamefont {Masaki}\ \bibnamefont
  {Shimizu}}\ and\ \bibinfo {author} {\bibfnamefont {Paul}\ \bibnamefont
  {Manneville}},\ }\bibfield  {title} {\enquote {\bibinfo {title} {Bifurcations
  to turbulence in transitional channel flow},}\ }\href@noop {} {\bibfield
  {journal} {\bibinfo  {journal} {Phys. Rev. Fluids}\ }\textbf {\bibinfo
  {volume} {4}},\ \bibinfo {pages} {113903} (\bibinfo {year}
  {2019})}\BibitemShut {NoStop}%
\bibitem [{\citenamefont {Xiong}\ \emph {et~al.}(2015)\citenamefont {Xiong},
  \citenamefont {Tao}, \citenamefont {Chen},\ and\ \citenamefont
  {Brandt}}]{Xiong15}%
  \BibitemOpen
  \bibfield  {author} {\bibinfo {author} {\bibfnamefont {Xiangming}\
  \bibnamefont {Xiong}}, \bibinfo {author} {\bibfnamefont {Jianjun}\
  \bibnamefont {Tao}}, \bibinfo {author} {\bibfnamefont {Shiyi}\ \bibnamefont
  {Chen}}, \ and\ \bibinfo {author} {\bibfnamefont {Luca}\ \bibnamefont
  {Brandt}},\ }\bibfield  {title} {\enquote {\bibinfo {title} {Turbulent bands
  in plane-poiseuille flow at moderate reynolds numbers},}\ }\href@noop {}
  {\bibfield  {journal} {\bibinfo  {journal} {Phys. Fluids}\ }\textbf {\bibinfo
  {volume} {27}},\ \bibinfo {pages} {84--468} (\bibinfo {year}
  {2015})}\BibitemShut {NoStop}%
\bibitem [{\citenamefont {Kanazawa}\ \emph {et~al.}(2017)\citenamefont
  {Kanazawa}, \citenamefont {Shimizu},\ and\ \citenamefont
  {Kawahara}}]{Kanazawa17}%
  \BibitemOpen
  \bibfield  {author} {\bibinfo {author} {\bibfnamefont {T.}~\bibnamefont
  {Kanazawa}}, \bibinfo {author} {\bibfnamefont {T.}~\bibnamefont {Shimizu}}, \
  and\ \bibinfo {author} {\bibfnamefont {G.}~\bibnamefont {Kawahara}},\
  }\bibfield  {title} {\enquote {\bibinfo {title} {Periodic solutions
  representing the origin of turbulent bands in channel flow},}\ }\href@noop {}
  {\bibfield  {journal} {\bibinfo  {journal} {Presented at KITP Conference:
  Recurrence, Self-Organization, and the Dynamics of Turbulence, 9 - 13
  January}\ } (\bibinfo {year} {2017})}\BibitemShut {NoStop}%
\bibitem [{\citenamefont {Paranjape}(2019)}]{Paranjape19}%
  \BibitemOpen
  \bibfield  {author} {\bibinfo {author} {\bibfnamefont {C.}~\bibnamefont
  {Paranjape}},\ }\bibfield  {title} {\enquote {\bibinfo {title} {Onset of
  turbulence in plane poiseuille flow},}\ }\href@noop {} {\bibfield  {journal}
  {\bibinfo  {journal} {Ph.D. Thesis, Institute of Science and Technology
  Austria, Klosterneuburg, Austria}\ } (\bibinfo {year} {2019})}\BibitemShut
  {NoStop}%
\bibitem [{\citenamefont {Xiao}\ and\ \citenamefont {Song}(2020)}]{Xiao20}%
  \BibitemOpen
  \bibfield  {author} {\bibinfo {author} {\bibfnamefont {Xiangkai}\
  \bibnamefont {Xiao}}\ and\ \bibinfo {author} {\bibfnamefont {Baofang}\
  \bibnamefont {Song}},\ }\bibfield  {title} {\enquote {\bibinfo {title} {The
  growth mechanism of turbulent bands in channel flow at low reynolds
  numbers},}\ }\href@noop {} {\bibfield  {journal} {\bibinfo  {journal} {J.
  Fluid Mech.}\ }\textbf {\bibinfo {volume} {883}},\ \bibinfo {pages} {R1}
  (\bibinfo {year} {2020})}\BibitemShut {NoStop}%
\bibitem [{\citenamefont {Tao}\ \emph {et~al.}(2018)\citenamefont {Tao},
  \citenamefont {Eckhardt},\ and\ \citenamefont {Xiong}}]{Tao18}%
  \BibitemOpen
  \bibfield  {author} {\bibinfo {author} {\bibfnamefont {Jianjun.}\
  \bibnamefont {Tao}}, \bibinfo {author} {\bibfnamefont {Bruno}\ \bibnamefont
  {Eckhardt}}, \ and\ \bibinfo {author} {\bibfnamefont {Xiangming.}\
  \bibnamefont {Xiong}},\ }\bibfield  {title} {\enquote {\bibinfo {title}
  {Extended localized structures and the onset of turbulence in channel
  flow},}\ }\href {\doibase 10.1103/PhysRevFluids.3.011902} {\bibfield
  {journal} {\bibinfo  {journal} {Phys. Rev. Fluids}\ }\textbf {\bibinfo
  {volume} {3}},\ \bibinfo {pages} {011902} (\bibinfo {year}
  {2018})}\BibitemShut {NoStop}%
\bibitem [{\citenamefont {Wang}\ \emph {et~al.}(2015)\citenamefont {Wang},
  \citenamefont {Li},\ and\ \citenamefont {E}}]{Wang15}%
  \BibitemOpen
  \bibfield  {author} {\bibinfo {author} {\bibfnamefont {Jianchun}\
  \bibnamefont {Wang}}, \bibinfo {author} {\bibfnamefont {Qianxiao}\
  \bibnamefont {Li}}, \ and\ \bibinfo {author} {\bibfnamefont {Weinan}\
  \bibnamefont {E}},\ }\bibfield  {title} {\enquote {\bibinfo {title} {Study of
  the instability of the poiseuille flow using a thermodynamic formalism},}\
  }\href@noop {} {\bibfield  {journal} {\bibinfo  {journal} {Proc. Natl. Acad.
  Sci.}\ }\textbf {\bibinfo {volume} {112}},\ \bibinfo {pages} {9518--9523}
  (\bibinfo {year} {2015})}\BibitemShut {NoStop}%
\bibitem [{\citenamefont {Mullin}(2011)}]{Mullin11}%
  \BibitemOpen
  \bibfield  {author} {\bibinfo {author} {\bibfnamefont {T.}~\bibnamefont
  {Mullin}},\ }\bibfield  {title} {\enquote {\bibinfo {title} {Experimental
  studies of transition to turbulence in a pipe},}\ }\href@noop {} {\bibfield
  {journal} {\bibinfo  {journal} {Annu. Rev. Fluid Mech.}\ }\textbf {\bibinfo
  {volume} {43}},\ \bibinfo {pages} {1--24} (\bibinfo {year}
  {2011})}\BibitemShut {NoStop}%
\bibitem [{\citenamefont {Chevalier}\ \emph {et~al.}(2007)\citenamefont
  {Chevalier}, \citenamefont {Schlatter}, \citenamefont {Lundbladh},\ and\
  \citenamefont {Henningson}}]{Chevalier2007}%
  \BibitemOpen
  \bibfield  {author} {\bibinfo {author} {\bibfnamefont {M.}~\bibnamefont
  {Chevalier}}, \bibinfo {author} {\bibfnamefont {P.}~\bibnamefont
  {Schlatter}}, \bibinfo {author} {\bibfnamefont {A.}~\bibnamefont
  {Lundbladh}}, \ and\ \bibinfo {author} {\bibfnamefont {D.~S.}\ \bibnamefont
  {Henningson}},\ }\bibfield  {title} {\enquote {\bibinfo {title} {{SIMSON - A}
  pseudo-spectral solver for incompressible boundary layer flows},}\
  }\href@noop {} {\bibfield  {journal} {\bibinfo  {journal} {Technical Report
  No. TRITA-MEK 2007:07}\ } (\bibinfo {year} {2007})}\BibitemShut {NoStop}%
\bibitem [{\citenamefont {Schmid}\ and\ \citenamefont
  {Henningson}(2001)}]{Schmid01}%
  \BibitemOpen
  \bibfield  {author} {\bibinfo {author} {\bibfnamefont {Peter~J.}\
  \bibnamefont {Schmid}}\ and\ \bibinfo {author} {\bibfnamefont {Dan~S.}\
  \bibnamefont {Henningson}},\ }\href@noop {} {\emph {\bibinfo {title}
  {Stability and Transition in Shear Flows}}}\ (\bibinfo  {publisher}
  {Springer, New York},\ \bibinfo {year} {2001})\BibitemShut {NoStop}%
\bibitem [{\citenamefont {Chomaz}\ \emph {et~al.}(1991)\citenamefont {Chomaz},
  \citenamefont {Huerre},\ and\ \citenamefont {Redekopp}}]{Chomaz91}%
  \BibitemOpen
  \bibfield  {author} {\bibinfo {author} {\bibfnamefont {J.-M.}\ \bibnamefont
  {Chomaz}}, \bibinfo {author} {\bibfnamefont {P.}~\bibnamefont {Huerre}}, \
  and\ \bibinfo {author} {\bibfnamefont {L.}~\bibnamefont {Redekopp}},\
  }\bibfield  {title} {\enquote {\bibinfo {title} {A frequency selection
  criterion in spatially developing flows},}\ }\href@noop {} {\bibfield
  {journal} {\bibinfo  {journal} {Stud. Appl. Maths}\ }\textbf {\bibinfo
  {volume} {84}},\ \bibinfo {pages} {119--144} (\bibinfo {year}
  {1991})}\BibitemShut {NoStop}%
\bibitem [{\citenamefont {Hammond}\ and\ \citenamefont
  {Redekopp}(1997)}]{Hammond97}%
  \BibitemOpen
  \bibfield  {author} {\bibinfo {author} {\bibfnamefont {D.}~\bibnamefont
  {Hammond}}\ and\ \bibinfo {author} {\bibfnamefont {L.}~\bibnamefont
  {Redekopp}},\ }\bibfield  {title} {\enquote {\bibinfo {title} {Global
  dynamics of symmetric and asymmetric wakes},}\ }\href@noop {} {\bibfield
  {journal} {\bibinfo  {journal} {J. Fluid Mech.}\ }\textbf {\bibinfo {volume}
  {331}},\ \bibinfo {pages} {231--260} (\bibinfo {year} {1997})}\BibitemShut
  {NoStop}%
\bibitem [{\citenamefont {Dee}\ and\ \citenamefont {Langer}(1983)}]{Dee83}%
  \BibitemOpen
  \bibfield  {author} {\bibinfo {author} {\bibfnamefont {G.}~\bibnamefont
  {Dee}}\ and\ \bibinfo {author} {\bibfnamefont {J.~S.}\ \bibnamefont
  {Langer}},\ }\bibfield  {title} {\enquote {\bibinfo {title} {Propagating
  pattern selection},}\ }\href@noop {} {\bibfield  {journal} {\bibinfo
  {journal} {Phys. Rev. Lett.}\ }\textbf {\bibinfo {volume} {50}},\ \bibinfo
  {pages} {383--386} (\bibinfo {year} {1983})}\BibitemShut {NoStop}%
\bibitem [{\citenamefont {Monkewitz}\ and\ \citenamefont
  {Nguyen}(1987)}]{Monkewitz87}%
  \BibitemOpen
  \bibfield  {author} {\bibinfo {author} {\bibfnamefont {P.}~\bibnamefont
  {Monkewitz}}\ and\ \bibinfo {author} {\bibfnamefont {L.}~\bibnamefont
  {Nguyen}},\ }\bibfield  {title} {\enquote {\bibinfo {title} {Absolute
  instability in the near-wake of two-dimensional bluff bodies},}\ }\href@noop
  {} {\bibfield  {journal} {\bibinfo  {journal} {J. Fluids Struct.}\ }\textbf
  {\bibinfo {volume} {1}},\ \bibinfo {pages} {165--184} (\bibinfo {year}
  {1987})}\BibitemShut {NoStop}%
\end{thebibliography}
%

\end{document}